\documentclass[preprint,superscriptaddress,preprintnumbers,amsmath,amssymb]{revtex4}
\usepackage{epsfig,graphicx}

\makeatletter
\def\@dotsep{4.5}
\makeatother

\begin{document}

\title{Anisotropic Collective Flow of $\Lambda$-Hyperons Produced in C + C Collisions at $4.2 \, \text{AGeV/$c$}$}

\author{L.\ Chkhaidze}
\email{ichkhaidze@yahoo.com} \email{ida@hepi.edu.ge}
\affiliation{Institute of High Energy Physics and Informatization, Tbilisi State University, Tbilisi, Georgia}
\author{P.\ Danielewicz}
\email{danielewicz@nscl.msu.edu}
\affiliation{National Superconducting Cyclotron Laboratory, Michigan State University, East Lansing, Michigan, USA}
\author{T.\ Djobava}
\email{Tamar.Djobava@cern.ch} \email{djobava@hepi.edu.ge}
\affiliation{Institute of High Energy Physics and Informatization, Tbilisi State University, Tbilisi, Georgia}
\author{L.\ Kharkhelauri}
\affiliation{Institute of High Energy Physics and Informatization, Tbilisi State University, Tbilisi, Georgia}
\author{E.\ Kladnitskaya}
\affiliation{Joint Institute for Nuclear Research, Dubna, Russia}

\begin{abstract}
Features of anisotropic collective flow and spectral temperatures have been determined for $\Lambda$ hyperons
emitted from C + C collisions, at incident momentum of $4.2 \, \text{AGeV/$c$}$, measured using the Propane Bubble
Chamber of JINR at Dubna.  Moreover, characteristics of protons and of negative pions, emitted from those
collisions, have been determined and provided for comparison.  The directed and elliptic flows of $\Lambda$s both
agree in sign with the corresponding flows of protons.
Parameters of the directed and elliptic flows for $\Lambda$s agree further, within errors, with the corresponding
parameters for the co-produced protons.  This contrasts an earlier finding by the E895 Collaboration of the directed flow being significantly weaker for $\Lambda$s than protons, in the much heavier Au + Au system, at comparable incident momentum.
Particle spectral temperatures in the C + C collisions have been determined focusing independently on either center-of-mass energy,
transverse energy or transverse momentum distributions.  For either protons or negative pions, the temperatures
were found to be approximately the same, no matter whether the emission of those particles was associated with
$\Lambda$ production or not.  Results of the measurements have been compared to the results of simulations
within the Quark-Gluon String Model.
\end{abstract}

\maketitle

\section{Introduction}

Produced strange particles represent important probes of high-density nuclear matter formed in relativistic
heavy-ion collisions.  Thus, because of the production thresholds, strangeness needs to be generated within
early stages of a collision.  Due to their longer mean free path, the strange particles are additionally likely
to leave the colliding system earlier than other hadrons.  Given the necessary contributions of the early
high-density stages of a collision, to the strangeness production, the yields of strange particles have been
linked theoretically to the nuclear equation of state of dense matter \cite{aich} and to the in-medium
modifications of particle masses \cite{weis}.  In drawing concrete conclusions on the equation of state from
yield data, comparison of the results from light C + C and heavy Au + Au systems turned out to be of crucial
importance~\cite{stur}.  Further in the literature, the~collective-flow characteristics of strange particles,
such as of $\Lambda$ hyperons to be investigated here, were shown~\cite{li1,li2} to be sensitive to the optical
potentials for those particles.  For $\Lambda$s, the flow characteristics were shown~\cite{li2,li3}, on the
other hand, to be relatively insensitive to the $\Lambda$-nucleon cross-sections.  In the beam-energy
of 2--10 AGeV, significant supranormal densities are reached in the collisions~\cite{dan}, accompanied by a good
degree of equilibration in heavy systems, conducive for studying the equation of state or optical potentials.
In this context, E895 Collaboration has measured yields and directed-flow characteristics of strange
particles~\cite{chung}, i.e.~$\Lambda$ hyperons and $K^0$ and $K_s^0$ mesons, in the semicentral Au + Au collisions
in the beam-energy range of 2--6 AGeV, at the AGS accelerator of the Brookhaven National Laboratory.
For $\Lambda$s, they have observed a lower flow than for protons, decreasing from $\sim 2/3$ to $\sim 1/3$ of the proton flow, across the 2--6 AGeV range.
The
experience \cite{stur} from the lower energy-range of 0.6--1.5 AGeV has clearly demonstrated, though, that results on
strangenesss from a light system, where a nontrivial equation of state might seem redundant, can be just as
important for learning about bulk nuclear properties, as results from a heavy system such as Au + Au.

In this paper, we present experimental results on characteristics of $\Lambda$ hyperons produced in the
light-system
C + C collisions  at incident momentum of 4.2 AGeV/$c$, registered in
the 2~m Propane Bubble Chamber (PBC-500) of JINR.
We concentrate on
anisotropic-flow characteristics and on spectral temperatures and compare our results for  $\Lambda$-hyperons
to those for protons and $\pi^-$ mesons from the C + C collisions.  We also examine spectral temperatures of
protons and $\pi^-$ mesons produced in association with $\Lambda$ hyperons.  The~anisotropic flows of protons
and of $\pi^-$ mesons have been studied before, on their own, in the C-induced collisions at
4.2--4.5 AGeV/$c$ in Dubna, specifically by the PBS-500 and the SKM-200-GIBS Collaborations
in the semicentral C + C collisions~\cite{chkha3} and in the
central C + Ne and C + Cu collisions \cite{chkha1, chkha2}.

In the next section, we address the details of our measurements.  Thereafter, we discuss the determination of the
mean reaction-plane component of $\Lambda$ transverse-momentum, as a~function of $\Lambda$ rapidity.  In
Section IV, we discuss the analysis of differential $\Lambda$ azimuthal-distribution and the determination
of $\Lambda$ elliptic flow parameter.  Temperatures of $\Lambda$'s, protons and $\pi^-$ mesons are addressed
in Section~V.  When presenting different experimental results, comparisons are made to the results of
Quark-Gluon String Model \cite{amel1, amel2}.  Our conclusions are presented in Section~6.

\section{
Experimental
Data}

For the experiment, the 2-m Propane Bubble Chamber (PBC-500) of
JINR has been placed in the magnetic field of 1.5~T.
Technical details behind the separation of C + C collisions in propane, identification of charged pions and
protons, application of different corrections and data processing, may be found in \cite{bond}.  Here, we
concentrate on the identification of $\Lambda$ hyperons.

The films from PBC-500 exposures have been scanned for $V^0$ events.  In a preliminary examination, the events
that either obviously represented $e^{+}$-$e^{-}$ pairs or did not point towards the nuclear interaction vertex
were rejected.  The~remaining events were measured in detail and reconstructed.  Specifically, the $V^0$ events
were tested, through fits, whether they fulfilled the kinematic conditions for $\Lambda$, $\bar{\Lambda}$,
$K_s^{0}$ decay or $\gamma$ conversion.  Finally, when a $V^0$ event was deemed to be the charged-particle decay,
$\Lambda \rightarrow p + \pi^{-}$, the momentum of the hyperon was reconstructed from decay products.  The
procedure resulted in 873 reconstructed $\Lambda$-hyperons.  For studying the collective flow of $\Lambda$s, an
exclusive analysis of the collision with an identified hyperon was required.

Depending on the analysis region, for a variety of reasons, some $\Lambda$ particles with charged decays could
not be identified.  Specifically, we estimate that about 26\% of the particles have been lost because their
decay occurred outside of the chamber effective region or too close, $l < 2 \, \text{cm}$, to the creation vertex.
Identification efficiency further deteriorated for certain azimuthal directions, leading to a loss of about 14\%
of the particles.  Finally, about 9\% of the particles have been lost in the forward c.m.\ hemisphere, for
directions too close to the beam.  Depending on the momentum region, corrections have been applied to the
identified particles, in the form of weights, compensating for the losses.  Additional information, on the
$\Lambda$-hyperon identification and on the corrections, can be found in Ref.~\cite{armut}.

For the further analysis, the C + C interactions have been selected, where, aside from a $\Lambda$, at least
4 participant protons have been identified.  In separating out the participant from fragmentation products,
the following criteria have been adopted: $p > 3.5 \, \text{GeV}/c$ and $\Theta < 3^\circ$ for projectile
fragmentation and $p < 0.25 \, \text{GeV}/c$ for target fragmentation.  It should be mentioned that events
with $\Lambda$s may contain an admixture of $K^+$ mesons; the average $K^+$ multiplicity is, in fact, half
of the $\Lambda$ multiplicity.  These are difficult to separate from proton and they contaminate the proton
sample, but on the average at the level of less than 10\%.

\section{Directed Flow of
$\Lambda$ Hyperons}

In determining the directed transverse flow of $\Lambda$ hyperons and of co-produced protons, we have employed
the method of
Danielewicz and Odyniec \cite{dan1}.
Most of the data below $4 \, \text{AGeV}$ in the literature have been, in fact, analyzed following that method.
The advantage of that method is that it can be employed even at small event statistics, such as typical
for film detectors.  The method relies on the summation over transverse momentum vectors of selected particles
in the events.

The reaction plane is spanned by the impact parameter vector $\vec{b}$ and the beam axis.
Within the transverse momentum method \cite{dan1}, the direction of $\vec{b}$ is estimated,
event-by-event, in terms of the vector $\vec{Q}$ constructed from particle transverse momenta
$\vec{p}_{i}^{\perp}$:
\begin{equation}\label{eq:Q}
  \vec{Q} = \sum_{i=1}^n \omega_i \vec{p}_{i}^{\perp} \, ,
\end{equation}
where the sum extends over all protons in an event.  The summation weight is
$\omega_i = 1$ for $y_i > y_c + \delta$, $\omega_i = -1$ for $y_i < y_c - \delta$ and
$\omega_i = 0$ for $y_c - \delta < y_i < y_c + \delta$, where $y_i$ is particle rapidity and
$y_c$ is the system c.m.\ rapidity.  Particles around the c.m.\ rapidity, with weak correlations with
the reaction plane, are not included in the reaction-plane determination.
In conducting the analysis in the c.m.\ system, we have $y_c =0$.  Otherwise, in suppressing the contribution
from midrapidity particles to the plane determination, we have employed $\delta=0.2$.

When referring a specific particle to the reaction plane, we exclude the contribution from that particle to the
estimate of the plane, to prevent an auto-correlation.  When referring a~proton, we exclude that proton; when referring
a lambda, we exclude the daughter proton.  The reference transverse vector for particle $j$ ($j > n$ for
$\Lambda$s and daughter protons) is then
\begin{equation}\label{eq:Qj}
\vec{Q}_j = {\sum_{\substack{i=1\\{i \ne j}}}^n} \omega_i \vec{p}_{i}^{\perp} \, ,
\end{equation}
where $i \ne j$ indicates exclusion of the aforementioned contribution and $n$ is the number of primary protons
identified in a collision.
Projection of the transverse momentum of particle~$j$ onto the estimated reaction plane
is
\begin{equation}\label{eq:pj}
p_j^{x \prime} = \frac{\vec{p}_j^{\perp}  \cdot \vec{Q}_j}{|\vec{Q}_j|} \, .
\end{equation}
The average in-plane momentum components $\langle p^{x \prime} (y)  \rangle $ can be obtained by averaging,
over events, the momenta in different rapidity intervals.

In a small system, such as ours, effects of transverse-momentum conservation in using~\eqref{eq:Qj} might be a source of concern~\cite{dan88,luk06}.  However, when the vector $\vec{Q}$ is dominated by particles with sizeable transverse momenta, that are detected with comparable efficiencies in the forward and backward hemispheres of a symmetric system, then the effects of conservation largely cancel between the hemishperes~\cite{dan88}.  A secondary effect is that of momentum conservation spreading out not only over initially present nucleons, but also produced pions.

Due to finite number of protons used in constructing the vector $\vec{Q}$ in (\ref{eq:Q}), the estimated reaction
plane fluctuates in the azimuth around the direction of the true reaction plane.  Because of those fluctuations,
the average momenta $\langle p^{x \prime}  \rangle$, calculated with the estimated reaction plane, get reduced as
compared to those
for the true reaction plane $\langle p^{x}  \rangle$:
\begin{equation}\label{eq:pxp}
\langle p^{x \prime} (y) \rangle = \langle p^{x} (y) \rangle  \, \langle \cos{\Phi} \rangle \, .
\end{equation}
Here, $\Phi$ is the angle between the true and estimated reaction plane.  The overall correction factor $k= 1/
\langle  \cos{\Phi} \rangle $, which needs to be applied to $\langle p^{x \prime} (y) \rangle$, in order
to obtain $\langle p^{x} (y) \rangle$, is subject to uncertainty, especially at low multiplicities.
In this work, we evaluate $ \langle \cos{\Phi} \rangle $ from the ratio \cite{dan1,beav}:
\begin{equation}\label{eq:cosfi}
\langle \cos{\Phi} \rangle = \frac{\langle \omega p^{x \prime} \rangle}{\langle \omega p^{x } \rangle}
= \left\langle \frac{\omega \, \vec{p}_j^{\perp} \cdot \vec{Q}_j}{| \vec{Q}_j |} \right\rangle
\left/ \left(  \frac{\left\langle Q^2 - \sum_{i=1}^n \left( \omega p_i^\perp  \right)^2 \right\rangle }{\langle n^2 - n \rangle}  \right)^{1/2} \right.
\, .
\end{equation}
For the ensemble of C + C collisions where a $\Lambda$ hyperon has been identified, we find $\langle \cos{ \Phi }
\rangle = 0.893$ and $k=1.12$.

Besides looking at the proximity of $\langle \cos{ \Phi } \rangle$ to~1, the accuracy of the reaction-plane
determination may be tested by applying the procedure
proposed in Ref.~\cite{dan1}.  Following that procedure, we have divided randomly each analyzed collision event
into two
sub-events and we have constructed vectors $\vec{Q}_1$ and $\vec{Q}_2$ for those sub-events.
The resulting distribution of reaction-plane directions from the sub-events peaks at zero degrees in relative
azimuthal angle and it exhibits a width of $\sigma \approx 50^\circ$.  This peaking directly testifies to the
presence of the special azimuthal direction in the events.
The distribution of reaction planes determined from full events, about the true reaction-plane direction, should
be about
half as wide as the relative distribution of directions from the sub-events~\cite{dan1,wils}.
The thus-estimated spread of the estimated reaction-plane directions about the true plane direction,
$\sigma_{0} \simeq \sigma/2 = 25^\circ$, is comparable
to the values reported
in \cite{jain} and to the values of $\sigma_{0}\approx 23 \div 25^\circ$ arrived at in Kr- and Au-induced
collisions with Ag(Br) targets \cite{adam}.

\begin{figure}
\includegraphics[width=.5\linewidth]{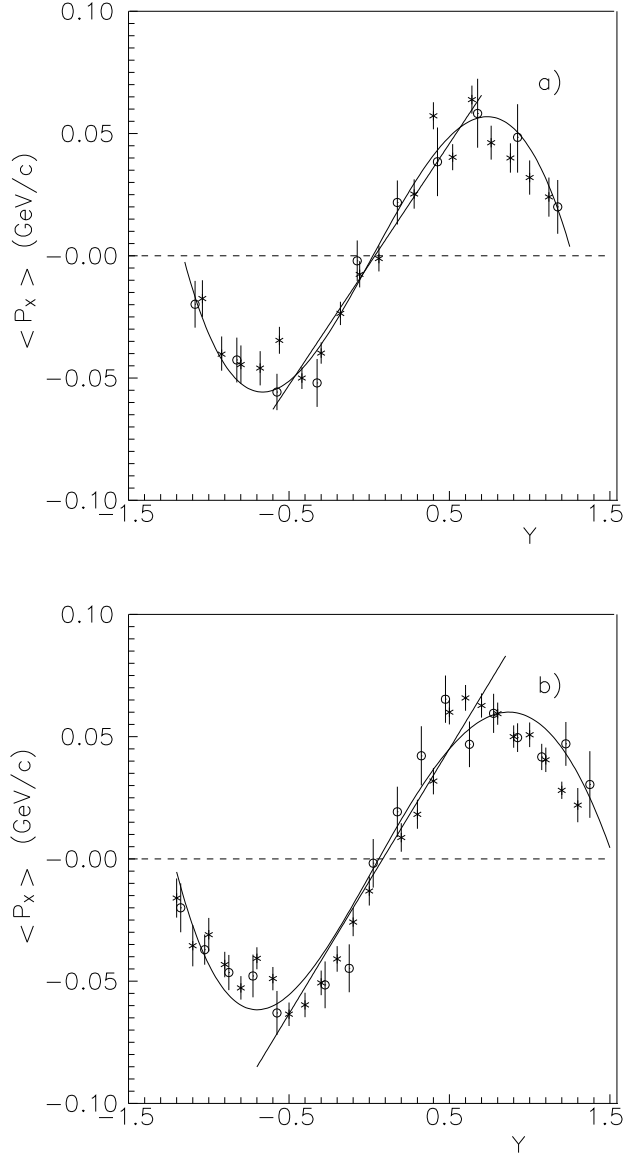}
\caption
{Average component of transverse momentum in the reaction plane, as a function of c.m.\ rapidity, for $\Lambda$
hyperons in panel (a), and for protons co-produced with the $\Lambda$s in panel (b).  The~data from central C + C
collisions at 4.2~AGeV/c, corrected
for $\langle \cos{\Phi} \rangle$, are represented by circles.  Results from simulations in the QGSM model are
represented by crosses.  The straight-line stretches represent
the slopes of the data at midrapidity cross-over and result from fitting the data with an odd third-order polynomial
within the rapidity region of $| y | < 0.65$ for (a), and $| y | < 0.70$ for~(b).  The curved lines guide
the eye over data.}
\label{fig:fig1}
\end{figure}

Figure~\ref{fig:fig1} shows the rapidity dependence of the average in-plane transverse-momentum component,
from Eq.~\eqref{eq:pj} and corrected for $\langle \cos{ \Phi } \rangle$, for $\Lambda$ hyperons in panel (a)
and for protons co-produced with $\Lambda$s in~(b).  It is apparent that the hyperons and protons flow in the
same in-plane directions, in the forward and backward hemispheres of C + C collisions at $4.2 \, \text{AGeV/$c$}$.
The dependencies $\langle p^{x}(y) \rangle$ are rather similar, even in detail, for protons and~$\Lambda$s,
with the magnitude of average momentum either similar for the two species or an edge higher for protons.

In the outer rapidity regions, the values of $\langle p^{x}(y) \rangle$ for protons may be affected by the
longer-term dynamics of nuclear spectator remnants, albeit rather reducing than enhancing proton momenta in
their magnitude.  One common measure, quantifying the dependence of~$ \langle p^{x} \rangle $ on rapidity, away
from the spectators, is the slope~\cite{doss} of
$ \langle p^{x} (y) \rangle $ at midrapidity:
\begin{equation}\label{eq:Fdef}
  F = \left. \frac{\text{d}\langle p^{x} \rangle}{\text{d}y} \right\vert_{y = 0} \, .
\end{equation}
To determine the slopes, marked with straight lines in the panels (a) and (b) of Fig.~\ref{fig:fig1}, odd
third-order polynomials have been fitted to the data in the midrapidity ranges of $| y | < 0.65$ for
$\Lambda$s, and $| y | < 0.75$ for protons co-produced with the $\Lambda$s.  The obtained flow parameters
are $F = 99 \pm 15 \, \text{MeV/$c$}$ for $\Lambda$s, and $F = 108 \pm 11 \, \text{MeV/$c$}$ for protons
co-produced with~$\Lambda$s.  The value of flow for all protons from the 10441 analyzed semi-central C + C
collision-events is $F = 113 \pm 10 \, \text{MeV/$c$}$.

Our C + C system is the lightest in the literature for which a directed flow of the $\Lambda$-hyperons has been
observed.  In spite of the system mass, the flow magnitude is similar to that observed for $\Lambda$s in other
systems within the specific energy regime.  Prior measurements of $\Lambda$ flow have been done by the E895
Collaboration at AGS~\cite{chung}, by the FOPI Collaboration at GSI~\cite{rit95}, and by the EOS Collaboration at LBL~\cite{just}.  In Au + Au collisions,
the E895 Collaboration obtained the $\Lambda$ flow of $F = 140 \pm 35$, $85 \pm 15$, and $65 \pm 5 \,
\text{MeV/}c$, at energies of 2, 4, and $6 \, \text{AGeV}$, respectively.  The~EOS Collaboration measured the
flow of $\Lambda$s of $F = 85 \pm 43 \, \text{MeV}/c$ in Ni + Cu collisions at~$2 \, \text{AGeV}$.  The E895
results indicate a decline of the $\Lambda$ flow on the high-end side of this energy regime.  This decline likely
reflects a reduction in time that the initial nuclei spend passing by each other.  Determination of $\langle p^x \rangle$ for $\Lambda$ hyperons emitted from Ni + Ni collisions at $1.93 \, \text{AGeV}$, by the FOPI Collaboration, has been done with a $p^\perp/m > 0.5$ cut on the $\Lambda$ transverse momenta.  Therefore, their results are more difficult to compare to those from other measurements.  However, within the FOPI data~\cite{rit95}, the results for $\Lambda$s and protons have been similar, after analogous cuts were applied to the particles.

Interestingly, while the directed flow is similar for $\Lambda$s in our C + C system and in the E895 Au + Au system at the comparable beam momentum of $4.8 \, \text{AGeV}/c$ at the energy of $4 \,\text{AGeV}$, this is not the case for protons, whether or not co-produced with the $\Lambda$s.  Thus, for all protons in Au + Au collisions, the E895 Collaboration has obtained~\cite{liu} flow parameter values of $F = 238 \pm 15$, $205 \pm 19$, and $183 \pm 16 \,
\text{MeV/}c$, at energies of 2, 4, and $6 \, \text{AGeV}$, respectively.  The E895 results for protons co-produced with $\Lambda$s in Au + Au \cite{chung} have been similar.  The Au + Au proton flow decreases across the energy range, but not as fast as the $\Lambda$ flow.  Importantly,
at the comparable incident momentum per nucleon to ours, of $4.8 \, \text{AGeV}/c$, the proton flow is about twice as high in Au + Au than in our C + C system, while the $\Lambda$ flow is similar.  This points to possible differences in the origin of the flow of protons and $\Lambda$ hyperons in the Au + Au and C + C systems, such as in a greater mean-field contribution to the proton flow in Au + Au.

Theoretically, the $\Lambda$ flow has been described in the past within the ARC cascade model~\cite{pang92} that
produced the value of $F = 113 \pm 31 \, \text{MeV}/c$ \cite{just} for the Ni + Cu collisions at $2 \,
\text{AGeV}$.  At the same time, the relativistic transport model RVUU~\cite{li1} produced the value of $F = 96
\pm 17 \, \text{MeV}/c$ for those collisions \cite{just}.  In spite of differences in the details of those models,
within the errors they both reproduce the $2 \, \text{AGeV}$ results.  Within the RVUU model, the $\Lambda$ mean-field potential has been fairly weak~\cite{li2}.  The RQMD model~\cite{sor95}, with no mean field for $\Lambda$s, produced $\Lambda$-flow that was consistent with E895 Au + Au data at $2 \, \text{AGeV}$ and underestimated the data at $4 \, \text{AGeV}$.

Our own measurements~\cite{chkha07} from the
collisions involving carbon, as either projectile or target, have been compared in the past to the predictions of the
Quark-Gluon String Model (QGSM).  This model combines string fragmentation with a hadronic cascade.
A~detailed description  of QGSM
can be found
in Refs.~\cite{amel1, amel2}.
Using the COLLI Monte-Carlo generator~\cite{amel3} based on QGSM, we have generated 7578 inelastic C + C collision
events, at the incident momentum of $4.2 \, \text{AGeV/$c$}$, where the production of $\Lambda$ hyperons has been
recorded.  When selecting participant protons, the same rejection criteria have been applied as to the protons in
the experiment.  In addition, from the analysis, the protons with deep angles, greater than $60^\circ$, have been
excluded, because experimental efficiency for detecting such steep tracks is greatly reduced.  In the analysis
of $\Lambda$ flow, only events with 4 or more participant protons were utilized. The average $\Lambda$ and
co-produced proton in-plane momenta are shown, together with the data, in Fig.~\ref{fig:fig1}.  It is apparent that
the model describes the data in a semi-quantitative manner.   Values of the flow parameter, determined within the
QGSM model, are $F = 97 \, \pm 8 \, \text{MeV}/c$ and $F = 108 \, \pm 8 \, \text{MeV}/c$, for the $\Lambda$s and for
co-produced protons, respectively.

\section{Elliptic Flow of $\Lambda$ Hyperons}

In the midrapidity region, where the directed flow is suppressed and the net particle abundance is the highest,
another form of emission anisotropy persists, of elliptic type.  It is associated with the geometry of the reacting
system in purely transverse directions.  The~overlap region of initial nuclei has an elliptic shape, and in this is more elongated
out of the reaction plane than in the plane.  The resulting pressure gradients are stronger in the in-plane than
out-of-plane directions, producing initially a stronger collective flow in the reaction plane direction than out of
the plane.  However, the spectator matter present in the vicinity of the reaction zone can suppress the growth of
collective motion within the plane, yielding in the end a stronger elliptic collective flow out of the plane,
commonly termed squeeze-out.

At high enough energies, the spectator matter moves away from the reaction zone quickly enough, though, to permit
the growth of in-plane flow, which then prevails.  Our incident energy is, actually, close to the energy where the
transition in the direction for the elliptic flow occurs.  Also, an elliptic emission pattern can be produced by
pure mean-free-path effects~\cite{heis99}.  With $\Lambda$s likely emitted over a shorter time than are emitted
protons, the~mean-free-path effects could produce a stronger out-of-plane elliptic pattern for $\Lambda$s than for protons.
In such a small system as ours, obviously, the mean-free-path effects have a greater chance to play a prominent
role than in a heavier system.

\begin{figure}
\includegraphics[width=.5\linewidth]{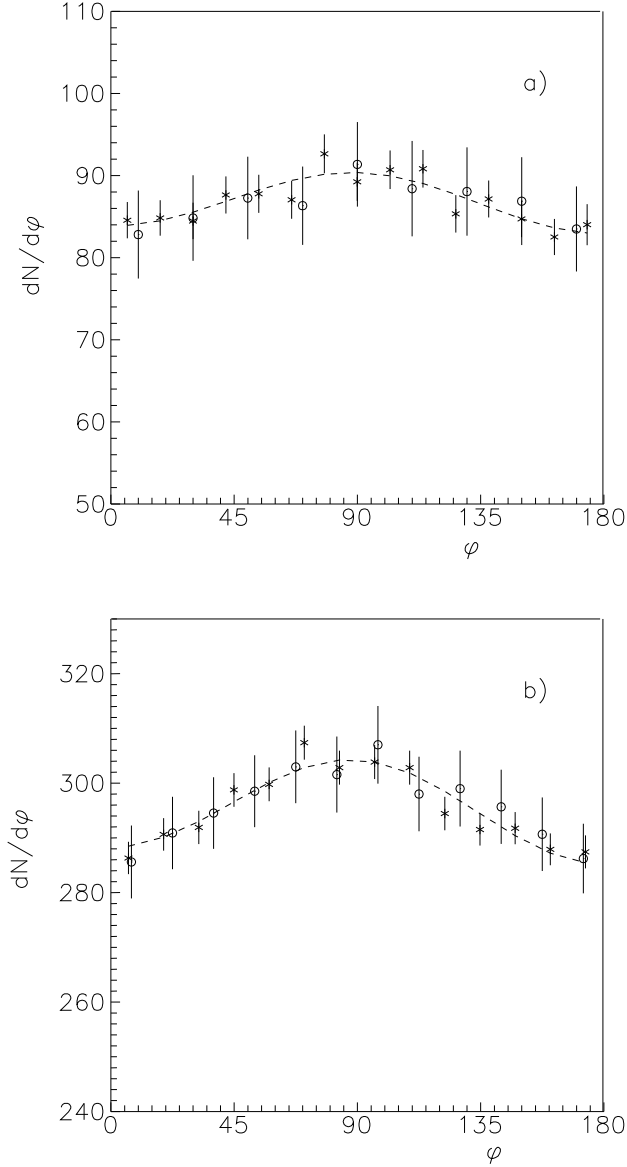}
\caption
{Distribution of midrapidity $\Lambda$s (a), and of co-produced protons (b),
in the azimuthal angle with respect to the estimated reaction plane,
in the C + C collisions at 4.2~AGeV/c.  The~$\Lambda$ hyperons stem from the rapidity region of $| y | < 0.65$ and the
protons stem from $| y | < 0.70$.  The circles represent data.  The crosses represent results of QGSM, with the
normalization set to match that of the data.  The lines represent fits to the data with the function $  {\text{d}
N}/{ \text{d} \varphi} = a_0 \, \big(  1 + a_1' \, \cos{\varphi} + a_2' \, \cos{2 \varphi} \big) $.
}
\label{fig:fig2}
\end{figure}

To gain an insight into the midrapidity emission pattern, we plot in Fig.~\ref{fig:fig2} the distribution of
$\Lambda$ hyperons (panel (a)), and of co-produced protons (panel (b)), in the azimuthal angle relative to the reaction
plane.  The angle for particle $j$ is obtained from $\cos{\varphi_j} = p_j^{x \prime}/ {p}_j^{ \perp}$ and we lump
together the left and right sides of the reaction plane, in order to increase statistics.  To~emphasize the
emission patterns, the ordinate axes do not start at zero in Fig.~\ref{fig:fig2}.

As is seen in Fig.~\ref{fig:fig2}, both the $\Lambda$ and co-produced proton distributions peak at 90$^\circ$,
i.e.~out of the reaction plane, exhibiting squeeze-out that is characteristic for lower incident energies, where
the spectator remnants modify the emission.  The two patterns in Fig.~\ref{fig:fig2} are actually quite similar;
approximately the two distributions differ just by an~overall normalization factor of about 3.35.  The distributions,
however, tend to be flattened by the fluctuations of the estimated reaction plane about the true reaction plane.
To better quantify the anisotropies, we fit the distributions with the function
\begin{equation}\label{eq:fco}
  \frac{\text{d} N}{ \text{d} \varphi} = a_0 \, \left(  1 + a_1' \, \cos{\varphi} + a_2' \, \cos{2 \varphi} \right)
\, .
\end{equation}
For squeeze-out, the coefficient $a_2'$ is negative.  Compared to the coefficient $a_2$, associated with a
distribution relative to the true reaction plane, $a_2'$ is reduced, though, by \cite{pink,andr}
\begin{equation}
\label{eq:a2red}
  a_2' = a_2 \, \, \langle \cos{2 \Phi} \rangle \, ,
\end{equation}
in an analogy to the reduction in (\ref{eq:pxp}).  The reduction coefficient for (\ref{eq:a2red}) may be estimated
from~\cite{chkha07}
\begin{equation}
\label{eq:c2f}
  \langle \cos{2 \Phi} \rangle = \frac{| \langle (p^{x \prime })^2 - ( p^{y \prime})^2 \rangle  |}{| \langle
(p^{x  })^2 - ( p^{y })^2 \rangle  |} \, ,
\end{equation}
where the numerator and denominator on the r.h.s.\ are, respectively, obtained from
\begin{equation}
  \langle (p^{x \prime })^2 - ( p^{y \prime})^2 \rangle = \bigg\langle 2 \bigg(
  \frac{\vec{p}_j^\perp \cdot \vec{Q}_j}{Q_j} \bigg)^2 - (p_j^\perp)^2 \bigg\rangle
  \, ,
\end{equation}
and
\begin{equation}
| \langle (p^{x })^2 - ( p^{y })^2 \rangle | =
\sqrt{ \frac{ \langle 2 \overline{\overline{T}} : \overline{\overline{T}} - \sum_{i=1}^n (p_i^\perp)^4
\rangle   }{\langle n^2 - n \rangle }   \, .        }
\end{equation}
In the above, the transverse tensor $\overline{\overline{T}}$ is
\begin{equation}
T^{\alpha \beta} = \sum_{i=1}^n \left( p_i^\alpha \,  p_i^\beta -
\frac{1}{2} (p_i^\perp)^2 \, \delta^{\alpha \beta} \right) \, , \hspace*{2em} \alpha=x,y \, ,
\end{equation}
and
\begin{equation}
\overline{\overline{T}} : \overline{\overline{T}} = \sum_{\alpha,\beta=x}^y T^{\alpha \beta} \, T^{\alpha \beta}
= \left( T^{xx} \right)^2 + \left( T^{yy} \right)^2 + 2 \left( T^{xy} \right)^2  \, .
\end{equation}

Use of Eq.~(\ref{eq:c2f}), for a reaction, requires that some elliptic anisotropy is present in the analyzed system
to start with, which, obviously though, is the case given the results in Fig.~\ref{fig:fig2}.
With Eq.~(\ref{eq:c2f}), using protons only, we find for our system $\langle \cos{2 \Phi} \rangle \simeq 0.581 $.
The fits to the azimuthal distributions with Eq.~(\ref{eq:fco}) are illustrated with solid lines in
Fig.~\ref{fig:fig2}.  Upon correcting the elliptical-modulation coefficients, following Eq.~(\ref{eq:a2red}),
we obtain $a_{2}= -0.062 \pm 0.031$  for $\Lambda$-hyperons and
$a_{2}= -0.049 \pm 0.018$ for the co-produced protons.  Within errors, those modulation patterns are similar.

We further compare our elliptic-flow results to the results from the QGSM model of the reaction.
The distributions relative to the reaction plane, obtained in the model following experimental procedures, and normalized according to the data,
are overlaid over the data in Fig.~\ref{fig:fig2}.   The~corresponding coefficient values are $a_{2}= -0.058 \pm 0.017$ for the $\Lambda$ hyperons (panel (a)) and $a_{2}= -0.047 \pm 0.008$ for co-produced
protons (panel~(b)).
As is apparent, the model predicts a clear signature of squeeze-out, the out-of-plane elliptic flow.  Further, within the experimental data, the model describes the elliptic-flow data.

Given the proximity of masses for protons and $\Lambda$-hyperons, the similarity of directed and elliptical flows
for $\Lambda$-hyperons and for associated protons could be principally understood in terms of a folding of local thermal
distribution with a field of collective velocity.  This kind of explanation normally assumes a local
grand-canonical ensemble with a universal temperature.  However, the production of strangeness could
tax the locally available energy and affect the universality of temperature, particularly in such a small
system as ours.  We next examine thermal characteristics of particle spectral distributions.

\section{Spectral Temperatures}

In the following, we shall examine the thermal characteristics of spectral distributions of $\Lambda$ hyperons
and of protons and negative pions either co-produced, or not, with the hyperons.  The distributions can
principally result from a superposition of local thermal distributions and collective velocity field at
freeze-out.  Obviously, pure thermal distributions cannot explain the asymmetries associated with the reaction
plane, that we have discussed.  However, here, we shall not attempt to separate those possible collective and
locally thermal contributions in a~folding.  Rather, we shall examine whether the effective temperatures, that could be associated
with the distributions, depend on species, on associated production and on the method of analysis of the
distribution.

Two primary types of collective motion may have a principal practical impact onto the spectral distributions.
Thus, an incomplete stopping of initial nuclei, or transparency, gives rise to a longitudinal collective motion.
In addition, collective expansion may develop in a reaction and be reflected in significant transverse radial flow.  In the following, we shall
examine both the spectral distributions in the net c.m.\ energy and in the transverse momentum or transverse mass.
The latter distributions are less likely to be affected by the longitudinal collective motion than the first.
Moreover, we shall explore the potentially different conclusions when confronting data with different versions
of the thermodynamic model, such as with the Hagedorn model, assuming particle freezeout within a net c.m.\ volume common for all particles, and with a model of thermal transverse distribution.
First, we analyze particle distributions in the net energy, in terms of the standard thermal model of freeze-out
with the c.m.\ system.

For a gas in thermal equilibrium within the volume $V$, the distribution in momentum of different species $i$ is
\cite{nag1, stock, kap}:
\begin{equation}
\begin{split}
\frac{\text{d}^{3}N_{i}}{\text{d}p^{3}} & =\frac{(2S_{i}+1)V}{(2\pi)^{3}}
\biggl[\exp{\Bigl(\frac{E_{i}-\mu_{i}}{T}\Bigr)}\pm 1\biggr]^{-1}
\\
 & \rightarrow \frac{(2S_{i}+1)V}{(2\pi)^{3}}
\exp{\Bigl(\frac{\mu_{i}-m_{i}}{T}\Bigr)} \exp{\Bigl(-\frac{E_{Ki}}{T}\Bigr)} \, ,
\end{split}
\end{equation}
where $V$ represents the freeze-out volume, $S_i$ is spin, $\mu_i$ is chemical potential, $T$ is temperature,
$E_i = \sqrt{m_i^2 + p^2}$ is the particle energy, and $E_{Ki} = E_i - m_i$ is the kinetic energy.  The $\pm$
signs refer to fermions and bosons, respectively, and the arrow indicates the Boltzmann-statistics limit of
$E_i - \mu_i \gg T$.  In a nondegenerate system, the distribution is then exponential in the kinetic energy,
with the temperature $T$ determining the slope of the exponential,
\begin{equation}
\frac{\text{d}^{3}N}{\text{d}p^{3}} = \frac{\text{d}^{2}N}{p^2 \, \text{d}p \, \text{d}\Omega}
= \frac{\text{d}^{2}N}{p  E \, \text{d}E \, \text{d}\Omega} =
\text{const} \cdot  \exp{\Bigl(-\frac{E_{K}}{T}\Bigr)} \, .
\end{equation}
Correspondingly, following the global freeze-out model, we can extract the freeze-out temperature by fitting an
exponential to the scaled distribution of particles in energy:
\begin{equation}
F(E_K) = (p E)^{-1} \, \text{d}N/\text{d}E_K = \text{const} \cdot \exp{(-E_K/T )}  \, .
\end{equation}

\begin{figure}
\includegraphics[width=1\linewidth]{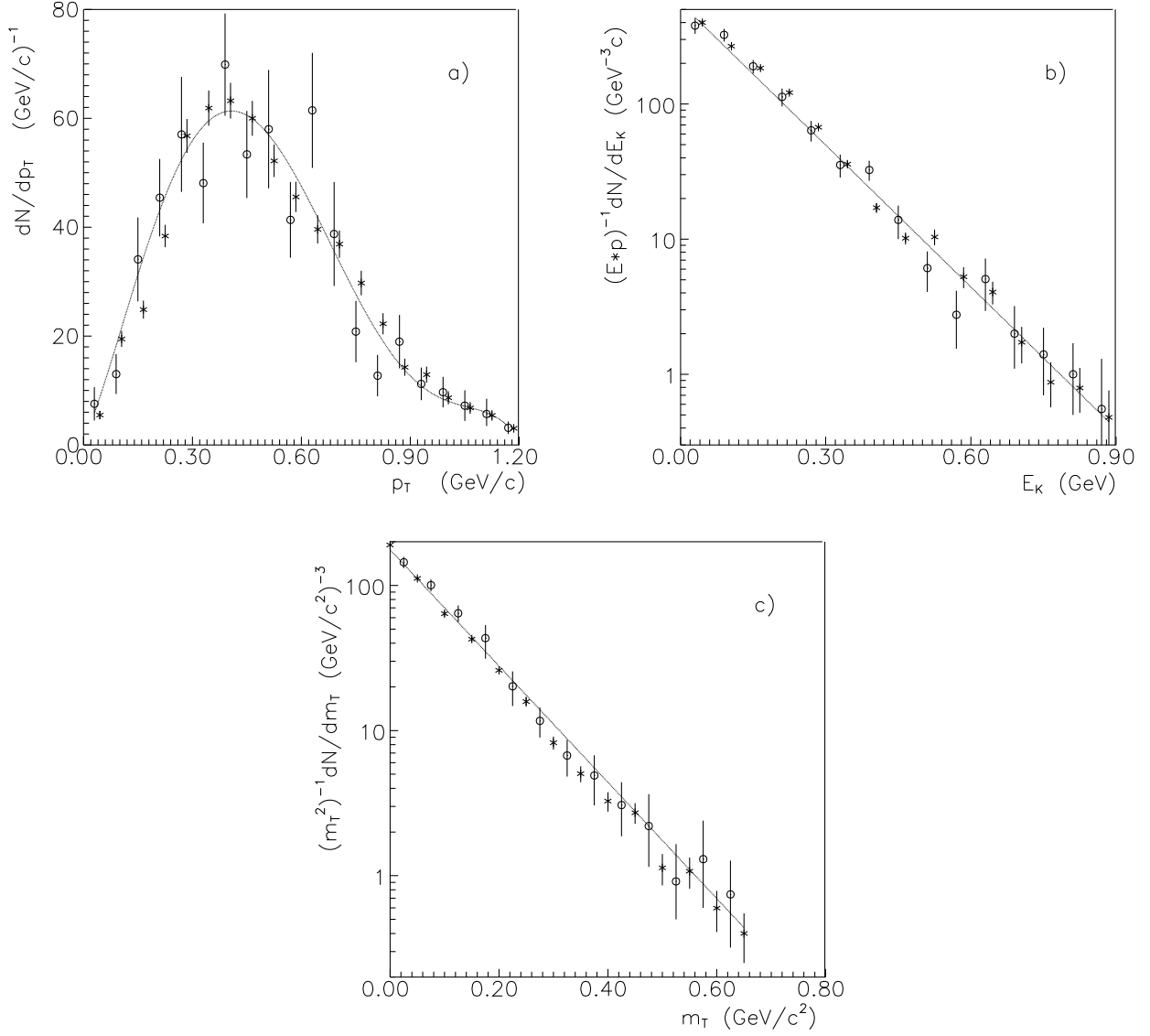}
\caption
{Spectra of $\Lambda$-hyperons emitted from C + C collisions, at $|y| < 0.6$: (a)~distribution of $\Lambda$s in the
transverse momentum, (b)~scaled distribution of $\Lambda$s in the c.m.\ kinetic energy, and (c)~scaled distribution
of $\Lambda$s in the transverse mass.  Circles represent data and stars represent QGSM results.  Lines represent
thermal fits to the data, discussed in the text.}
\label{fig:fig3}
\end{figure}

\begin{figure}
\includegraphics[width=1.\linewidth]{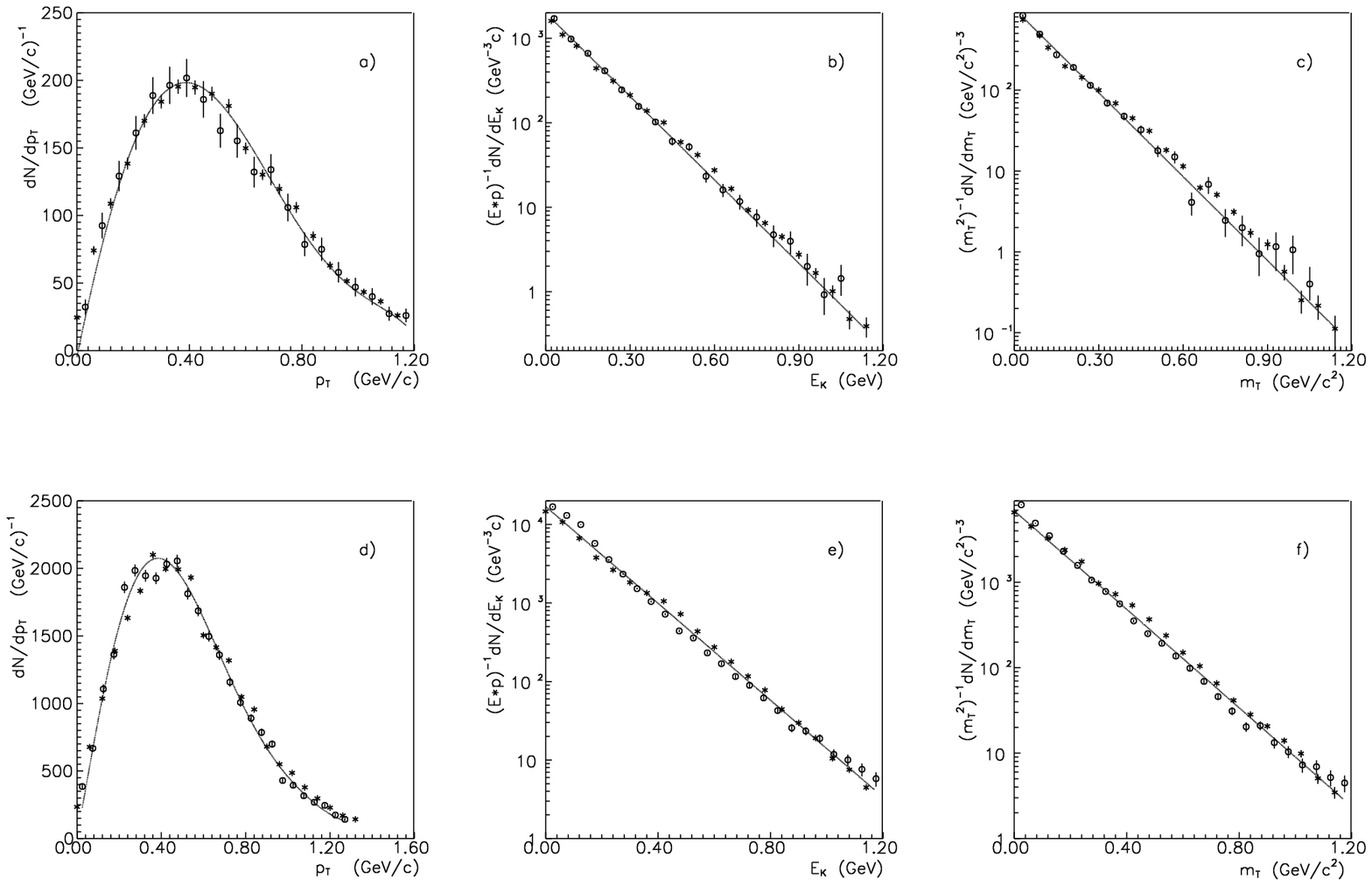}
\caption
{Spectra of protons produced in the presence (a,b,c) or absence (d,e,f) of $\Lambda$ hyperons, at $|y | < 0.6 $. The leftmost panels
(a,d) show the distribution of protons in the transverse momentum, the~center panels (b,e) show the scaled distribution of protons in the c.m.\ kinetic
energy, and the rightmost panels (c,f) show the scaled distribution of protons in the transverse mass.  Circles represent data and stars represent
QGSM results.  Lines represent thermal fits to the data, discussed in the text.}
\label{fig:fig4}
\end{figure}

\begin{figure}
\includegraphics[width=1.\linewidth]{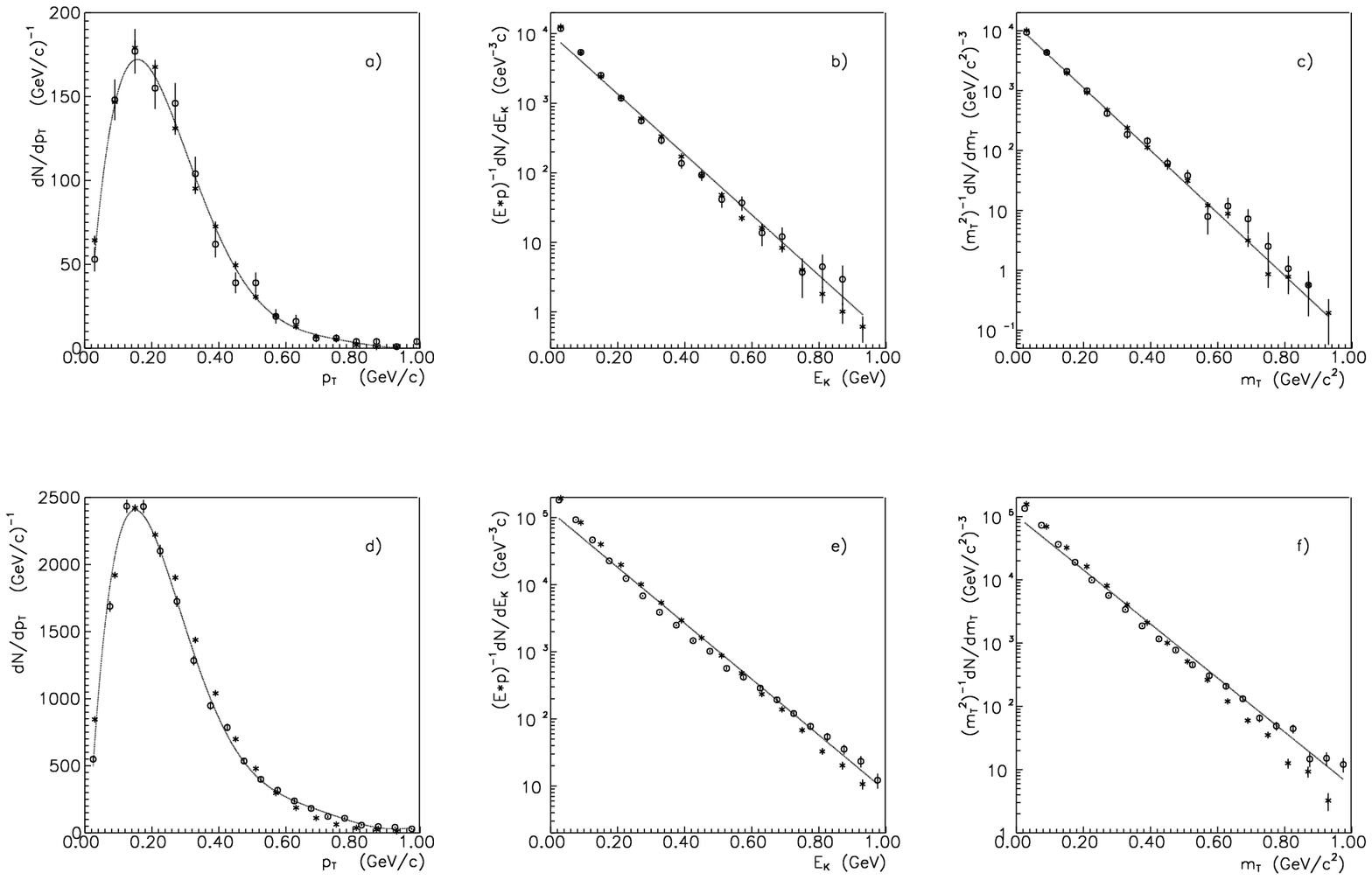}
\caption
{Spectra of $\pi^-$ mesons produced in the presence (a,b,c) or absence (d,e,f) of $\Lambda$ hyperons, at~$|y | < 0.8$.  The leftmost panels (a,d) show the distributions of mesons in the transverse momentum, the~center panels (b,e) show the scaled distributions of mesons
in the c.m.\ kinetic energy, and the rightmost panels (c,f) show the scaled distributions of mesons in the transverse mass.  Circles represent
data and stars represent QGSM results.  Lines represent thermal fits to the data, discussed in the~text.}
\label{fig:fig5}
\end{figure}

Figure~\ref{fig:fig3} shows the scaled distribution $F$ for $\Lambda$s in the rapidity region of $|y| < 0.6$.
In~addition, Figs.~\ref{fig:fig4} and~\ref{fig:fig5} show, respectively, the distribution for protons at
$|y| < 0.6$ and $\pi^-$~mesons at $|y| < 0.8 $, either emitted in the presence or absence of $\Lambda$s.  In each case, the
scaled distributions are quite well described by an exponential. The only significant deviations are observed
for the lowest-energy $\pi^-$ mesons and these deviations likely combine the Coulomb effects and effects of
baryon-resonance decays.  Temperatures from fits to the kinetic-energy spectra are next provided in
Table~\ref{tab:info}.  The temperatures are similar, but not the same, for different species.  For $\Lambda$s, we
find the temperature of $T = 111 \pm 7 \, \text{MeV}$, which is higher than the temperature for $\pi^-$ mesons of the
order of $85 \, \text{MeV}$, and is lower than the temperature for protons of the order of~$134 \, \text{MeV}$.  A lower
spectral temperature for pions than for baryons might be associated with collective motion that has more effect on
the more massive baryons than on pions.  The~difference between protons and $\Lambda$s could be principally due to
the effects of transparency, that should be of greater importance for the protons than for hyperons produced centrally.  In
examining, in particular, this last possibility, we next turn to transverse spectra.

The Hagedorn Thermodynamic Model~\cite{hag1,hag2} allows, principally, for a set of fireballs displaced from each
other in rapidity.  In that model, particles with different momenta freeze-out within a volume that is of universal
magnitude when assessed in the rest-frame for any given momentum.  In that model, the distribution in transverse
momentum is of the shape
\begin{equation}
\label{eq:Hagedorn}
\text{d}N/\text{d}p^{\perp} = \text{const} \cdot p^{\perp} \, m^{\perp} \, \text{K}_\text{1}(m^{\perp}/T)\simeq
\text{const}
\cdot p^{\perp} \, (m^{\perp} \, T)^{1/2} \, \exp{(-m^{\perp}/T)} \, ,
\end{equation}
where $\text{K}_\text{1}$ is the MacDonald function~\cite{grad}, the transverse mass
is $m^\perp = \sqrt{m^2 + p^{\perp 2}} $ and the approximation above is valid for $m^\perp \gg T$. For
large $m^\perp$ in different thermal models, the~scaled distribution in transverse mass is exponential in the
mass:
\begin{equation}
\label{eq:scaledmt}
F_\alpha(m^{\perp})= (m^{\perp})^{-\alpha} \, \text{d}N/\text{d}m^{\perp}= \text{const}
\cdot \exp{(-m^{\perp}/T)} \, ,
\end{equation}
with the Hagedorn model of Eq.~(\ref{eq:Hagedorn}) being one example where $\alpha = 3/2$.  In the following, we
shall fit the transverse spectra either using the Hagedorn model or following Eq.~(\ref{eq:scaledmt}) with
$\alpha=2$, in order to explore model sensitivity.  We will display those fits, respectively, in the context of
spectra presented as distributions in transverse momentum, Eq.~(\ref{eq:Hagedorn}), or scaled distribution in
transverse mass, Eq.~(\ref{eq:scaledmt}).  The two distributions emphasize different parts of the spectrum.

In different panels, Figs.~\ref{fig:fig3}, \ref{fig:fig4} and \ref{fig:fig5} present the transverse distributions,
and the aforementioned fits to those distributions, respectively, for $\Lambda$-hyperons and for protons and $\pi^-$ mesons, either
produced in the presence or absence of $\Lambda$s.  It is apparent that thermal descriptions work also well for
the transverse spectra.  The temperatures deduced from transverse spectra are provided in Table~\ref{tab:info}.
The temperatures for baryons are similar to those deduced from distributions in the c.m.\ kinetic energy,
contradicting the supposition that the latter distributions may be significantly affected by transparency.
Greater differences in deduced temperature values are found for the $\pi^-$ mesons.  This may be due to the
combination of nonthermal features of spectra, rapidity cut, and nearly ultrarelativistic nature of pions in the collisions.
The regularities which are observed in analyzing the distributions, irrespectively of the spectrum or the thermal
model, are that the $\pi^-$-meson temperatures are lower than the proton temperatures and that the $\Lambda$
temperatures are lower than the proton temperatures.

As we have already indicated, the lower $\pi^-$ temperatures may be attributed to the presence of a collective
motion, which has a stronger effect on the more massive protons than on pions.  Given the moderate azimuthal
asymmetries, as in Fig.~\ref{fig:fig2}, that collective motion must be predominantly azimuthally symmetric, since
a significant motion needs to underly the significantly different temperatures.  The origin of different
temperatures for protons and $\Lambda$s is likely tied to the rest-energy cost involved in strangeness production.

Thus, because of the rest-energy cost involved in strangeness production, the strange particles must typically
emerge, from their production process, at low velocities relative to their environment.  Subsequent interactions
may broaden the hyperon spectrum, but not enough to make its width comparable to that of proton spectrum, if the hyperons
decouple from the system soon after their production.  An early freeze-out for the hyperons may be facilitated by the
lower interaction cross-sections with the environment, for the hyperons, than for other baryons.  Because of the
early freeze-out, the kinetic energies of hyperons might not tap much, in addition, the energy reservoir
represented by the excess rest-energy in baryonic resonance excitations.  During system expansion, that excess
energy gets converted into the radial collective energy, but some of that conversion may take place only after
the $\Lambda$s have decoupled from the system. The rest-energy involved in strangeness production taxes also
energies of other particles.  In a small system such as C + C, some shrinking of spectra can occur and a~hint of
this may be seen in Table \ref{tab:info} in that the temperatures for protons and pions are marginally, but
systematically, lower in the presence than in the absence of a~hyperon, irrespectively of the thermodynamic model
employed.

The QGSM has the capability of describing the discussed effects, associated with strangeness energy cost and
with interaction cross sections.  The results for spectra from the QGSM model are compared to data in
Figs.~\ref{fig:fig3}, \ref{fig:fig4} and~\ref{fig:fig5}.  The temperatures from fits to the QGSM spectra are
further compared to those for data in Table~\ref{tab:info}.  It is apparent that the model describes adequately
the measured spectra.  In addition, the model reproduces variation of the temperatures with species and with thermodynamic
models fitted to different spectra, and even reproduces the marginal decrease of temperature for associated production.

Elsewhere in the literature, differences in the temperatures of $\Lambda$s and protons have been also detected.
Thus, in Ni + Cu collisions at $2 \, \text{AGeV}$, measured at the Bevalac, the EOS Collaboration~\cite{just}
has arrived at the $\Lambda$ temperature of $106 \pm 5 \, \text{MeV}$ and proton temperature of $142 \pm 1 \,
\text{MeV}$, in purely thermal fits to $m^\perp$-spectra.  Analogous fits to spectra from the ARC
model~\cite{pang92}, for those collisions, gave rise~\cite{just} to the temperatures of $91 \pm 2 \, \text{MeV}$
and $121 \pm 1 \, \text{MeV}$, respectively, for $\Lambda$s and protons.  In Ni + Ni collisions at $1.93 \,
\text{AGeV}$, measured at GSI SIS, the FOPI Collaboration~\cite{mer08} has arrived at the temperatures of about
$119 \, \text{MeV}$ and $139 \, \text{MeV}$, when fitting $m^\perp$-spectra of $\Lambda$s and of protons,
respectively, around midrapidity.

An attempt to separate the collective from thermal motion, for $\Lambda$s emitted from central Au + Au collisions
at $11.6 \, \text{AGeV/$c$}$, has been carried out by the E891 Collaboration~\cite{ahmad} who have fitted their
measured $\Lambda$-spectra in terms of an anisotropic blast-wave model.  Their fit has produced the temperature
of $96 \pm 37 \, \text{MeV}$ in combination with an average radial velocity of ($0.40 \pm 0.12 \,) c$, over the laboratory
rapidity range of $(1.7 - 3.2)$.  Within errors, those spectral parameters are consistent with those needed to
describe the spectra of light nuclei from a very similar central system of Au + Pb
at $11.5 \, \text{AGeV/$c$}$, measured by the E864 Collaboration~\cite{arm00}.  Putting the large errors aside, one
should mention that, for a~large number of interactions following strangeness production in a large system,
there is no principal reason for the $\Lambda$ and proton temperatures to be very different.

From the measurements of $\Lambda$s and of associated yields,
for which temperatures got determined,
in the same
incident-momentum range,
we should finally mention the Dubna measurements~\cite{avram} of Mg + Mg collisions at $4.3 \,
\text{AGeV/$c$}$.  The temperatures have been determined there following the Hagedorn model.  Rather than from fits
to the spectra, however, those temperatures have been inferred from average transverse momenta.  For negative pions,
the temperatures of $90 \pm 1 \, \text{MeV}$ and $92 \pm 2 \, \text{MeV}$ have been deduced, with and without
$\Lambda$s, respectively.  For $\Lambda$s, the temperature of $137 \pm 9 \, \text{MeV}$ has been deduced.  The
pion temperatures are comparable to those in our measurements.  On the other hand, the $\Lambda$ temperature is
significantly higher than ours.  Principally, a somewhat stronger radial collective motion in a larger system
could be responsible for some increase in temperature for baryons.

\section{Summary}

In this paper, we have assessed the characteristics of anisotropic collective flow and spectral temperatures of
$\Lambda$ hyperons emitted from the central C + C collisions at $4.2 \, \text{AGeV}/c$, measured using the Propane
Bubble Chamber of JINR at Dubna.  In addition, characteristics of protons and of negative pions emitted from those
collisions have been determined and provided for comparison.

Directed particle flow has been assessed following the transverse-momentum method of Danielewicz and Odyniec.
The $\Lambda$ hyperons have been found to flow in the same direction as protons and, otherwise, to exhibit the
rapidity dependence of in-plane transverse momentum that is quantitatively close to that for protons.  The
midrapidity flow parameter $F$ was found to take on the value of $99 \pm 15 \, \text{MeV}/c$ for $\Lambda$s
and $108 \pm 11 \, \text{MeV}/c$ for co-produced protons.  The similarity of the directed flows for $\Lambda$s and co-produced protons contrasts earlier findings of the E895 Collaboration for the Au + Au system at comparable incident momentum, where the $\Lambda$ flow has been found at least twice as strong as the proton flow.

Elliptic flow in the C + C system, studied around midrapidity, turned out, as the directed flow,
to be of the same sign for $\Lambda$s and protons and was found to point in the squeeze-out direction, out of the reaction
plane.  Within errors, the~elliptic flow parameters $a_2$ are consistent for those particles, at $-0.062 \pm 0.031$
for $\Lambda$s and $-0.049 \pm 0.018$ for the co-produced protons.

The spectral temperatures have been determined for $\Lambda$s and protons and negative pions produced either in the
absence or presence of the $\Lambda$-hyperons.  Both the c.m.-energy and the transverse energy and momentum spectra
have been examined, as well as different variants of the thermodynamic model used to fit data.  Lower temperatures
have been arrived at for $\Lambda$-hyperons and for negative pions than for protons.  In addition, the negative-pion
temperatures have been found to be a bit lower than the $\Lambda$ temperatures.  A hint of lowering of proton and pion
temperatures in the presence of $\Lambda$ hyperons has been detected.

Experimental results have been compared to the results of collision simulations within the Quark-Gluon String Model
which combines string decays with a hadronic interaction cascade.  The model describes the observations rather well.

\begin{acknowledgements}
The authors are very grateful to \framebox{N.Amelin} for providing them with access to the QGSM code
program COLLI. One of the authors (L.Ch.) would like to thank the board of Directors of the
Laboratory of High Energies  of JINR for the warm hospitality and also
thank J.~Lukstins and O.~Rogachevsky for assistance during the manuscript preparation.
This work was partially supported by the U.S.\
National Science Foundation under Grant Nos.\ PHY-0555893 and PHY-0800026 and by the Georgian National Science
Foundation under Grant GNSF/ST08/4-418.

\end{acknowledgements}

\listoftables

\listoffigures

\newpage

\begin{table}
\caption{The temperatures of $\Lambda$-hyperons, and of protons and
$\pi^{-}$ mesons, produced either in the presence or absence of $\Lambda$s, inferred from different spectra following different thermal-model assumptions
discussed in the text.\\[-1.5ex]}
\label{tab:info}
\begin{tabular}{|c|c|c|c|c|c|}    \hline
          &  Spectrum     & Number     & \multicolumn{3}{c|}{Spectral Temperature} \\
Particles &  Origin       &   of       & \multicolumn{3}{c|}{  $T$~(MeV) for     }  \\
          &               & Particles  &
\multicolumn{1}{c|}{ d$N/\text{d}p^{\perp} $}&
\multicolumn{1}{c|}{ d$N/\text{d}E_{K} $}&
\multicolumn{1}{c|}{ d$N/\text{d}m^{\perp} $}\\
\hline
$\Lambda$ & Data & 561 &
 $105 \pm 7$ & $111 \pm 7$ &  $103 \pm 6$ \\
\cline{2-6}
$|y| < 0.6 $ & QGSM &7578 &
 $106 \pm 3$ & $105 \pm 3$ &  $102 \pm 3$ \\
\hline
$p$ with $\Lambda$  & Data & 3961 &
 $135 \pm 5$ & $132 \pm 4$ &  $128 \pm 4$ \\
\cline{2-6}
$|y| < 0.6 $  & QGSM & 33653 &
 $133 \pm 3$ & $128 \pm 3$ &  $129 \pm 3$ \\
\hline
$p$, no $\Lambda$ & Data &76002 &
  $140 \pm 3$ & $136 \pm 3$ &  $134 \pm 3$ \\
\cline{2-6}
$|y| < 0.6 $  & QGSM &213560 &
 $136 \pm 2$ & $133 \pm 2$ &  $130 \pm 2$ \\
\hline
$\pi^-$ with $\Lambda$ & Data & 1146 &
 $94 \pm 2$ & $85 \pm 2$ &  $81 \pm 2$ \\
\cline{2-6}
$|y| < 0.8 $  & QGSM &12005 &
 $93 \pm 2$ & $88 \pm 2$ & $81 \pm 2$ \\
\hline
$\pi^-$, no $\Lambda$ & Data & 23781 &
 $98 \pm 2$ & $86 \pm 2$ &  $85 \pm 2$ \\
\cline{2-6}
$|y| < 0.8 $ & QGSM &66820 &
 $96 \pm 2$ & $90 \pm 2$ &  $84 \pm 2$ \\
\hline
\end{tabular}
\end{table}


\end{document}